\documentclass[twoside,twocolumn]{article}

\usepackage{abstract} 
\usepackage{amsmath} 
\usepackage[english]{babel} 
\usepackage{booktabs}
\usepackage[hang, small,labelfont=bf,up,textfont=it,up]{caption} 
\usepackage{enumitem} 
\usepackage{fancyhdr}
\usepackage[T1]{fontenc}
\usepackage[hmarginratio=1:1,top=32mm,columnsep=20pt]{geometry} 
\usepackage{graphicx} 
\graphicspath{ {./images/} }

\usepackage{hyperref} 
\usepackage{listings, lstautogobble} 
\usepackage[sc]{mathpazo} 
\usepackage{microtype} 
\usepackage{titlesec} 
\usepackage{titling}
\usepackage{url} 
\usepackage{xcolor} 

\renewcommand\thesection{\Roman{section}} 
\renewcommand\thesubsection{\roman{subsection}} 

\setlist[itemize]{noitemsep} 

\titleformat{\section}[block]{\large\scshape\centering}{\thesection.}{1em}{} 
\titleformat{\subsection}[block]{\large}{\thesubsection.}{1em}{} 

\title{\huge DFTWS for blockchain: Deterministic, Fair and Transparent Winner Selection} 
\author{%
\textsc{Felix Hoffmann, Udo Kebschull} \\[1ex] 
\normalsize FIAS, Goethe University Frankfurt am Main \\ 
\normalsize \href{mailto:felix.hoffmann@iri.uni-frankfurt.de}{felix.hoffmann@iri.uni-frankfurt.de, uk@rz.uni-frankfurt.de} 
}
\date{\today} 
\renewcommand{\maketitlehookd}{
\begin{abstract}
\noindent
  This publication describes the block winner selection process that will be used in a novel Proof-of-Useful-Work blockchain for High Energy Physics that the authors are currently working on. Instead of spamming hashing operations to mine blocks, miners will be running Monte Carlo simulations to support a real-world HEP experiment with useful data. The block problems will be defined by a Root Authority which is represented by a HEP experiment like CBM. The focus in this publication is a mechanism that allows the Root Authority to select a winner from a list of nodes that solved a block problem. The mechanism is designed so that winner selection is deterministic, fair and transparent. This mechanism allows every node to verify the fairness of the winner selection process without giving the nodes a tool to be able to improve their own winning chances.
\end{abstract}
}

\begin{document}
\maketitle

\section{Introduction}
The authors of this publication are working on a Proof-of-Useful-Work blockchain for High Energy Physics, in which new blocks are created by running Monte Carlo simulations that support a real-world HEP experiment while at the same time securing the underlying blockchain instead of wasting energy by spamming hashing operations as common in traditional Proof-of-Work that still is being used in e.g. Bitcoin \cite{bitcoin}. In order to generate \textit{useful} data, there must be an entity that can make use of the resulting data. As a result, HEP experiment insiders (who are able to determine which kind of Monte Carlo simulations are needed) act as a Root Authority that defines problems that miners need to solve. This publication focuses on the winner selection process of such as blockchain.\\\\
Each node generates an Ed25519 \cite{ed25519} private key of length 32 bytes. Then the corresponding public key of length 32 is derived. The public key is then used to derive a libp2p \textit{node ID} that identifies the node. Messages sent between nodes are signed using Ed25519 signatures on Curve25519.\\\\
In order to create a new block, the Root Authority (e.g. a member of the CBM collaboration \cite{cbm}) defines the parameters for a Monte Carlo simulation that must be run in a provided HEP environment that every miner runs locally. The RNG of the Monte Carlo simulations is derived block the previous block hash and is part of the problem definition. As a result, the resulting ROOT (common file format in HEP) files are identical for each node that runs the same simulation (reproducibility). This enables a probabilistic solution verification mechanism that is not the focus of this paper. There will never be the exact same (sub) problems chosen for two different blocks, because the results would be identical, and it would give an advantage to miners who participated in the blockchain longer than others and have a memory of past solutions.\\
Each problem must be solved within a defined interval of time (part of the problem definition). If a node has solved the block problem, it uploads the resulting data to RA servers (only RA has read access here) and publicly broadcasts a zero knowledge proof (Keccak-256 \cite{keccak} of solution hash) that retrospectively shows that this node was able to solve the problem in time without revealing the solution hash itself. This enables the following mechanism:
\begin{itemize}
	\item A node can prove that it solved the problem in time. Otherwise, it would not have been able to determine the correct solution hash, hash it again and then publicly broadcast publish signed solution hash and hashed solution hash.
	\item Nodes can not steal the solutions of other nodes because they can see $Keccak256(Keccak256(solution data))$ of other nodes, but they need to sign $Keccak256(solution data)$ which they can not derive from $Keccak256(Keccak256(solution data))$.
	\item When the RA has chosen a winner, a list of all nodes that solved the problem and the correct solution hash are also included in the block. This allows each node to verify that the message the other nodes signed was in fact $Keccak256(solution data)$. As a result, nodes can control each other (correct solution was published) and nodes can control the RA: The list of nodes that solved the problem can be verified because with the knowledge of the correct solution hash it can be verified that the message the other nodes signed and broadcast in time was the hash of the correct solution. If a node signed the correct solution but is not included in the winner list, then each node can instantly detect that the RA can not be trusted. Furthermore, nodes are able to determine that the most commonly broadcast solution hash has been chosen as correct solution by the RA (this is useful in combination with the probabilistic solution determination algorithm which is not the focus of this paper).
	
\end{itemize}
It should be noted that from the list of nodes that published a correct solution only one winner is chosen. This is done to prevent token inflation (winner gets a block reward token that is halved in regular block intervals) and to prevent nodes from performing a Sybil attack to calculate the solution once and then collect the reward for multiple identities they control.

\section{Winner selection}
The question that will be answered in this section is: How can one winner be chosen from the list in a fair and transparent way?
\subsection{Procedure}
Before problem is broadcast:
\begin{enumerate}
	\item Root authority (RA) generates 10000 pseudo-random bytes. 10000 can be replaced with any large value, as its only purpose is to prevent brute-force guessing. This value is used so that nodes who do not know it can not in real-time calculate who currently would be chosen as winner.
	\item RA calculates $s = Keccak256(PreviousBlockHash + random bytes)$ ["+" means concatenation], and adds $s$ to the block problem definition (along with all the simulation settings miners need to run the simulations). This means value $s$ is broadcast along with the problem, but the input that led to $s$ will only be revealed by the RA after the problem validity interval runs out.
\end{enumerate}
As soon as problem validity interval ran out and no further solutions are accepted. At this point in time, a list of all nodes (nodeID + signature) that submitted the correct solution is known. More specifically, each node had broadcast its Ed25519 signature of $Keccak256(solutionHash)$ before uploading the solution to RA servers. All these signatures are put into a list by the RA to create e.g. the following simplified list of two nodes that submitted the correct solution. Note that the list is ordered by NodeID alphabetically ascending:
\begin{table}[h!]
	\centering
	\begin{tabular}{||c c||}
		\hline
		NodeID & Signature \\ [0.5ex]
		\hline\hline
		12D3KooWaaa & abc \\
		12D3KooWbbb & def \\ 
		12D3KooWccc & acf \\ [1ex]
		\hline
	\end{tabular}
\end{table}
\\The task is to choose one winner from this list:
\begin{enumerate}
	\setcounter{enumi}{2}
	\item Publish random bytes that were used to calculate $s$. Now every miner can verify $s = Keccak256(PreviousBlockHash + random bytes)$.
	\item Concatenate all signatures from the list into $c$. The list must be ordered alphabetically ascending so that every node can reproduce this procedure. 
	\item Calculate $a = Keccak256(c+randombytes)$ and take the first 15 characters (so that it later can be represented by a 64 bit integer). With this, there now exists a value that every blockchain participant can reproduce, but no one could have predicted this value before the problem solution was due: The RA knew the random bytes, but can not possibly predict the resulting problem solution signatures of any node as this value can only be created by a node that knows the solution hash and its own private key. There is no way for the RA to choose the random bytes in a specific way to influence the winner selection outcome, otherwise Keccak-256 would need to be replaced with a different cryptographically secure hashing algorithm. A node knows its own private key but not any private keys of other nodes, which means that there is no way to predict the resulting signatures of other nodes. Additionally, the random bytes the RA used to create the known value $s$ will not be known until the point in time where problem solution submissions are not allowed anymore and the ordering of the resulting list is agreed to be alphabetically ascending, so there not only is no way for a node to influence its position on the resulting list (NodeID can not be changed due to signature mechanism) but there also is no way to influence the resulting problem solution signature (Ed25519 signatures are deterministic given same private key and data to sign).hex
	\item Represent the hex string $a$ with its decimal representation and calculate $winnerindex = a \mod len(list)$. The resulting index will be the winner index in the list. Even if nodes were able to influence the position on this list, they would not be able to calculate any index that is preferred over another index. Every blockchain participant is able to calculate $winnerindex$ after the RA publishes the required information which means that there is consensus on who should be selected block winner of this block.
\end{enumerate}
The procedure described above is deterministic (every node can follow it and reach the same results), fair (there is no reason to prefer any list position over another list position and the RA can not influence the winner selection itself in some preferred direction) and transparent (the winner selection procedure is public knowledge and can be reproduced by any blockchain participant).\\
If the RA breaks protocol every node is able to notice it. The RA can not collude with a subset of nodes (by leaking the at this point in time secret $randombytes$) because the resulting signatures of the nodes that are not colluding can not be known in time (if at least one node is not colluding there is no way to determine the winnerindex beforehand). An additional security measure is that the random bytes are combined with the previous block hash, which even if there were a vulnerability, would limit the time for any pre-calculations as the hash of the previous block can not be known before a specific point in time. Every node knows that the generation of random bytes happens at a point in time where the at a later time resulting signatures are not known, so there is no reason to think that there exists a malicious choice of random bytes. A node is not able to fake signatures and the list cannot contain nodes that had broadcast invalid signatures (invalid as in the data that was signed was not the solution hash which now is known by everyone and which is the same for everyone).\\
The data that is broadcast by nodes while solution submissions are still allowed does not in any way help to determine the solution of the problem, which means a node has to solve the block problem to get added to the winner selection list. Faking node identities (and therefore improving winning chances by having multiple entries on the list) in a Sybil attack will not be possible, because in order to join the blockchain a node must generate a private key, derive its public key, derive its nodeID and then register its real-world identity along with its nodeID at the Root Authority. As long as the real-world identity registration process is transparent for all blockchain participants, there is no reason to question the fairness of the winner selection process described in this paper.

\subsection{Implementation in Golang}
The code on the next page implements the idea described in this paper for a simplified scenario. The following code randomly generates five libp2p node identities, lets them sign an imaginary solution hash using Ed25519 and then shows how to select a winner among the nodes.
\onecolumn 
\begin{lstlisting}[language=go]
// tested with go 1.20.6
package main

import (
  "fmt"			// stdout
  "sort"		// sorting
  "strings"		// alphabetically sorting
  "strconv"		// hex -> decimal
  "encoding/hex"	// BytesToString/StringToBytes
  "github.com/libp2p/go-libp2p/core/crypto"// ed25519
  "golang.org/x/crypto/sha3"		   // keccak256
  libp2p "github.com/libp2p/go-libp2p"	   // libp2p nodeID
)

type ExampleNode struct {
  PrivKey		crypto.PrivKey
  PubKey		crypto.PubKey
  NodeID		string		// libp2p nodeID
  ExampleSig		string		// signature
}

func main() {
  // example solution hash
  solHashString := "69868b59cab0f269284b96acca
5549ab804095fcb452d64aba3c904bc82117bc"

  // example previous blockhash
  // prevBlockHash := "d60ee5d9b1a312631632d0ab
  //                   8816ca64259093d8ab0b4d29f35db6a6151b0f8d"

  // example random bytes 
  // (shorter length and constant to keep example simple)
  randBytes := "a4896a3f93bf4bf58378e579f3cf193bb4af
1022af7d2089f37d8bae7157b85f"

  // Root Authority publishes
  // Keccak256(prevBlockHash + randBytes) and then later
  // reveals randBytes

  // hold example nodes in slice
  nodeList := []ExampleNode{}

  // generate  5 nodes for this example
  for c:=0; c < 5; c+=1 {
    //	generate keys
    priv, pub, err := crypto.GenerateKeyPair(
      crypto.Ed25519, // generate ed25519 keys
	-1,             // use default key len
    )
    if err != nil {
	  panic(err)
    }

    // get libp2p nodeID
    h, err := libp2p.New(
      libp2p.Identity(priv),
    )
    if err != nil {
	  panic(err)
    }
    nodeID := h.ID().String()
	
    // get []byte representation of solHash
    solHash, err := hex.DecodeString(solHashString)
    if err != nil {
	panic(err)
    }
	
    // sign it
    sig, err := priv.Sign(solHash)	// returns []byte
    if err != nil {
	panic(err)
    }
	
    // ensure signature is valid
    verifiedSigSuccess, err := pub.Verify(solHash,sig)
    if err != nil {
	panic(err)
    }
    if !verifiedSigSuccess {
	panic("Invalid signature!")
    }
	
    // get string representation of sig
    sigString := hex.EncodeToString(sig)
	
    // create example node object
    n := ExampleNode {
	  PrivKey:	priv,
	  PubKey:	pub,
	  NodeID:	nodeID,
	  ExampleSig:	sigString,
    }
    
    
    // append this node to slice of nodes
    nodeList = append(nodeList, n)

  } // end of for loop

  // sort list of nodes alphabetically ascending
  // (0<9<a<z and case-insensitive) by nodeID
  sort.Slice(nodeList, func(i, j int) bool {
      return strings.ToLower(nodeList[i].NodeID) < 
             strings.ToLower(nodeList[j].NodeID)
    }
  )

  // print all sorted nodeIDs and their signatures
  for i, node := range nodeList {
    fmt.Printf("Node %v: %v \nSignature: %v \n\n", i, 
    node.NodeID, 
    node.ExampleSig)
  }

  // Event:  Problem not valid anymore, no further solution
  // submissions allowed. Root Authority publishes
  // values solHashString and randBytes. This means nodes 
  // can now verify that other nodes signed the correct 
  // solution hash and they can follow 
  // the following winner selection procedure because 
  // randBytes is now publicly known. The list of winners 
  // in this example is nodeList

  // ---- Winner selection ----

  // winner list is sorted already. now concatenate all 
  // signatures into c
  c := ""
  for _, curNode := range nodeList {
    c += curNode.ExampleSig
  }

  // calculate a = Keccak256(c + randomBytes)
  a := Keccak256(c + randBytes)

  // take first 15 chars of a and convert to decimal number
  // calculate winner_index = a mod len(winnerlist)
  aDec, err := strconv.ParseInt(a[:15], 16, 64) 
  if err != nil {
    panic(err)
  }

  // calculate winner_index = a mod len(winnerlist)
  winnerIndex := aDec % int64(len(nodeList))

  // print nodeID of winner
  fmt.Printf("----\nWinner Index: %v \nBlock Winner: %v \n", 
    winnerIndex, 
    nodeList[winnerIndex].NodeID)
  }

// Keccak256 takes string and returns its Keccak256 hash as string
func Keccak256(input string) string {
  // you could replace below with New256() but then 
  // it will not show same output as common 
  // online comparison tools 
  hash := sha3.NewLegacyKeccak256()
  hash.Write([]byte(input))
  hashSum := hash.Sum(nil)
  hashString := hex.EncodeToString(hashSum)

  return hashString
}

\end{lstlisting}

\begin{center}\label{fig:code}
	\includegraphics[scale=0.56]{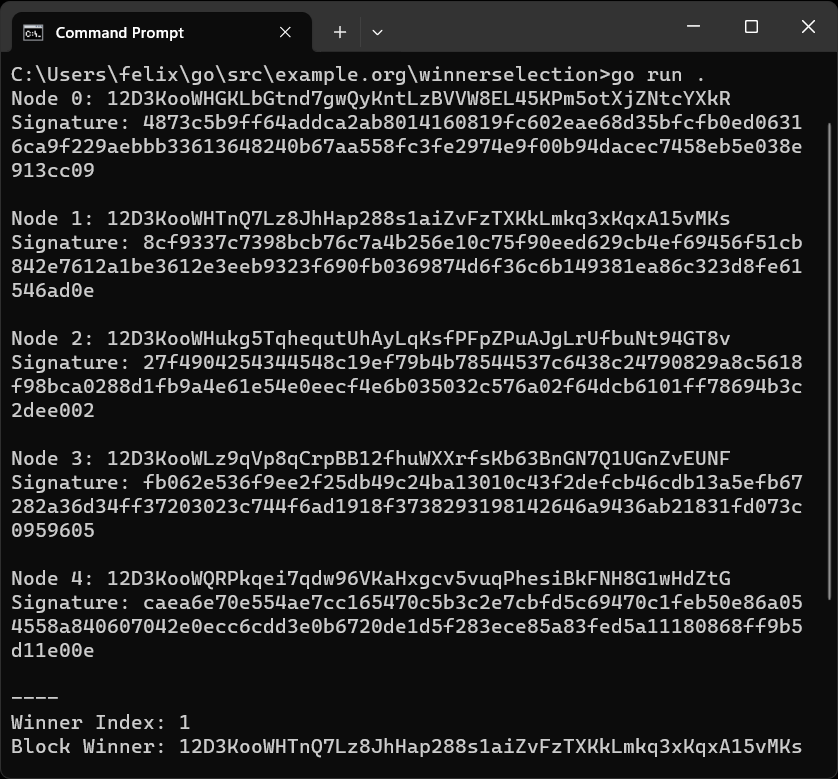}
\end{center}
\bibliographystyle{alpha}
\bibliography{references}

\end{document}